\documentclass[12pt]{iopart}
\usepackage{amssymb,setstack}

\def\ba{\begin{array}}
\def\ea{\end{array}}
\def\be{\begin{equation}}
\def\ee{\end{equation}}
\def\bea{\begin{eqnarray}}
\def\eea{\end{eqnarray}}
\def\nn{\nonumber}
\newtheorem{veta}{Theorem}
\newtheorem{lemma}{Lemma}

\def\pd{\partial}
\def\sp{{\rm span}}

\def\C{{\mathbb C}}
\def\R{{\mathbb R}}
\def\N{{\mathbb N}}
\def\F{{\rm F}}
\def\a{{\mathfrak a}}
\def\f{{\mathfrak f}}
\def\g{{\mathfrak g}}
\def\h{{\mathfrak h}}
\def\l{{\mathfrak l}}
\def\n{{\mathfrak n}}
\def\r{{\mathfrak r}}
\def\s{{\mathfrak s}}
\def\sl{{\mathfrak sl}}
\def\t{{\mathfrak t}}
\def\z{{\mathfrak z}}

\newcommand{\sdir}{\ensuremath{\rlap{\raisebox{0.15ex}{$\mskip 6.5mu\scriptstyle+$}}\supset}}

\begin{document}

\title{A class of solvable Lie algebras and their Casimir Invariants}

\author{L \v Snobl\dag  and P Winternitz\ddag}

\address{\dag\ Centre de recherches math\'ematiques, Universit\'e de Montr\'eal, CP 6128, Succ Centre-Ville, Montr\'eal 
(Qu\'ebec) H3C 3J7, Canada and 
Faculty of Nuclear Sciences and Physical Engineering, 
Czech Technical University, B\v rehov\'a 7, 115 19 Prague 1, Czech Republic }

\address{\ddag\ Centre de recherches math\'ematiques and Departement de math\'ematiques et de statistique, 
Universit\'e de Montr\'eal, CP 6128, Succ Centre-Ville, Montr\'eal (Qu\'ebec) H3C 3J7, Canada}

\eads{\mailto{Libor.Snobl@fjfi.cvut.cz}, \mailto{wintern@crm.umontreal.ca}}

\begin{abstract}
A nilpotent Lie algebra $\n_{n,1}$ with an $(n-1)$ dimensional Abelian ideal is studied. All indecomposable solvable Lie 
algebras with $\n_{n,1}$ as their nilradical are obtained. Their dimension is at most $n+2$. The generalized Casimir invariants of 
$\n_{n,1}$ and of its solvable extensions are calculated. For $n=4$ these algebras figure in the Petrov classification of 
Einstein spaces. For larger values of $n$ they can be used in a more general classification of Riemannian manifolds.
\end{abstract}

\pacs{02.20.Qs,03.65.Fd,04.50.+h}

\section{Introduction}\label{intro}

The purpose of this article is to classify a certain type of finite dimensional solvable Lie algebras, existing for any dimension 
$n$ with $n \geq 4$.
These Lie algebras will be described below.
Here we shall first present our motivation for performing this investigation.

Lie groups and Lie algebras appear in physics in many different guises. They may be a priori parts of the physical theory, 
like Lorentz or Galilei invariance of most theories, or the (semi)simple  Lie groups of the Standard model in particle theory.

Alternatively, specific Lie groups may appear as consequences of specific dynamics. Consider any physical system with dynamics 
described by a system of ordinary or partial differential equations. This system of equations will be invariant under some 
local Lie group of local point transformations, taking solutions into solutions. This symmetry group $G$ and its Lie algebra $\g$
can be determined in an algorithmic manner \cite{Olver}. The Lie algebra $\g$ is obtained as an algebra of vector fields, usually
in some nonstandard basis, depending on the way in which the algorithm is applied.

An immediate task is to identify the algebra found as being isomorphic to some known abstract Lie algebra. To do this we must 
transform it to a canonical basis in which all basis independent properties are manifest. Thus, if $\g$ is decomposable into 
a direct sum, it should be explicitly decomposed into components that are further indecomposable
\be\label{direct}
\g = \g_1 \oplus \g_2 \oplus \ldots  \oplus \g_k.
\ee
Each indecomposable component must be further identified. Let $\g$ now denote such an indecomposable Lie algebra.
A fundamental theorem due to E. E. Levi \cite{Levi,Jac}, tells us that any finite--dimensional Lie algebra can be represented as the
semidirect sum
\be\label{semidirect}
\g = \l \sdir \r, \;  [\l,\l] = \l, \ [\r,\r] \subset \r, [\l,\r] \subseteq \r,
\ee
where $\l$ is semisimple and $\r$ is the radical of $\g$, i.e. its maximal solvable ideal. If $\g$ is simple, we have $\r \sim 0$. 
If $\g$ is solvable, we have $\l \sim 0$.

Semisimple Lie algebras over the field of complex numbers $\C$ have been completely classified by Cartan \cite{Cartan}, over 
the field of real numbers $\R$ by Gantmacher \cite{Gan} (see e.g. \cite{Hel}).

Algorithms realizing decompositions (\ref{direct}),(\ref{semidirect}) exist \cite{Rand}. The ``weak'' link in the classification of
Lie algebras is that not all solvable Lie algebras are known.

There are two ways of proceeding in the classification of Lie algebras, in particular solvable ones: by dimension, or by structure.

The dimensional approach for real Lie algebras was started by Bianchi \cite{Bianchi} 
who classified all real Lie algebras of dimension 2 and 3.
Those of dimension 4 were classified by Kruchkovich \cite{Kru1}.
Further work in this direction is due to Morozov (nilpotent
Lie algebras up to dimension 6) \cite{Morozov}, Mubarakzyanov \cite{Mub1,Mub2,Mub3,Mub4}, 
Patera et al. \cite{PSW} and Turkowski
\cite{Tur1,Tur2}. The classification of low--dimensional 
Lie algebras over $\C$ was started earlier by S. Lie himself \cite{Lie}.

The most interesting physical application of the classification of low--dimensional Lie algebras is in general relativity. Indeed, 
the classification of Einstein spaces according to their isometry groups \cite{Petrov} is based on the work of Bianchi 
and his successors \cite{Bianchi,Kru1}.
The Petrov classification concerns Einstein spaces of dimension 4 and hence involves isometry groups of relatively low dimensions 
\cite{Petrov,SKMHH}. 

String theory \cite{GSW,Johnson}, brane cosmology \cite{RS} and 
some other elementary particle theories going beyond the standard model require the use of higher dimensional spaces. 
Any attempt at a Lie group classification of such spaces will require knowledge of higher--dimensional Lie groups, including 
solvable ones.

It seems to be neither feasible, nor fruitful to proceed by dimension in the classification of Lie algebras $\g$ beyond 
${\rm dim} \ \g=6$. It is however possible to proceed by structure. 

Any solvable Lie algebra $\g$ has a uniquely defined nilradical ${\rm NR}(\g)$, i.e. maximal nilpotent ideal, satisfying
\be\label{dimnilrad}
{\rm dim} \ {\rm NR}(\g) \geq \frac{1}{2} \ {\rm dim} \ \g.
\ee
Hence we can consider a given nilpotent algebra of dimension $n$ as a nilradical and then find all of its extensions to solvable 
Lie algebras. In previous articles this has been performed for the following nilpotent Lie algebras:
Heisenberg algebras $\h(N)$ (where $\dim \h(N)=2N+1, \ N \geq 1$) \cite{RW}, Abelian Lie algebras $\a_n, \ n \geq 1$ \cite{NW,NW1},
``triangular'' Lie algebras $\t(N), \ (\dim \t(N) =\frac{N(N-1)}{2}, \ N \geq 2 )$ \cite{TW,TW1}.

Here we shall consider a class of nilpotent algebras that, for want of a better notation, we shall call $\n_{n,1}$,
 where the subscript denotes the dimension of $\n_{n,1}$, $n=3,4,\ldots$
This algebra has an $(n-1)$ dimensional Abelian ideal with the basis $( e_1,\ldots,e_{n-1} )$. The Lie brackets are given by
\bea
\nn [e_j,e_k] & = & 0, \; 1 \leq j, k \leq n-1, \\
\nn [e_1,e_n] & = & 0,\\
\label{nla} [ e_k,e_n] & = & e_{k-1}, \; 2\leq k \leq n-1.
\eea
Thus the action of the element $e_n$ on the Abelian ideal is given by an indecomposable nilpotent Jordan matrix
\be
M = \left( \ba{ccccc} 0 & 0 &  \ldots & 0 & 0 \\
1 & 0 &  \ldots & 0 & 0 \\
0 & 1 &  \ldots & 0 & 0 \\
\vdots  &  & \ddots &   & \vdots \\
0 & 0  & \ldots & 1 & 0
\ea \right) \in \F^{(n-1) \times (n-1)}
\ee
We shall consider this algebra over the field $\F$, where we have $\F=\R$, or $\F=\C$.

We mention that for $n=3$ we have $\n_{3,1}\simeq \h(1) \simeq \t(3)$. 
The algebra $\n_{4,1}$ is the only 4--dimensional indecomposable nilpotent Lie algebra. The algebra $\n_{n,1}$ exists for 
any integer $n$ satisfying $n \geq 3$ and for $n \geq 4$ it is no longer isomorphic to $\h(N)$ nor $\t(N)$.

\section{Mathematical preliminaries}

\subsection{Basic concepts}

Three different series of subalgebras can be associated with any given Lie algebra. The dimensions of the subalgebras in each of these 
series are important characteristics of the given Lie algebra.

The {\it derived series} 
$ \g = \g^{(0)} \supseteq \g^{(1)} \supseteq \ldots \supseteq \g^{(k)} \supseteq \ldots $ 
is defined recursively
\[ \g^{(k)} = [\g^{(k-1)},\g^{(k-1)}], \ \g^{(0)}=\g .\]
If the derived series terminates, i.e. there exists $k \in \N$ such that $\g^{(k)} \sim 0$, then $\g$ is called a 
{\it solvable Lie algebra}.

The {\it lower central series} $ \g = \g^{0} \supseteq \g^{1} \supseteq \ldots \supseteq \g^{k} \supseteq \ldots $ 
is again defined recursively
\[ \g^{k} = [\g^{k-1},\g], \; \g^{0}=\g. \]
If the lower central series terminates, i.e. there exists $k \in \N$ such that $\g^{k} \sim 0$, then $\g$ is 
called a {\it nilpotent Lie algebra}. The lowest value of $k$ for which we have $\g^{k} \sim 0$ is the degree of 
nilpotency of a nilpotent Lie algebra.

Obviously, a nilpotent Lie algebra is also solvable. An Abelian Lie algebra is nilpotent of degree 1.

The {\it upper central series} is $ \z_{1} \subseteq \ldots \subseteq \z_{k} \subseteq \ldots \subseteq \g$. 
In this series $\z_1$ is the {\it center} of $\g$
\[ \z_1 = C(\g) = \{ x \in \g | [x,y]=0, \ \forall y \in \g \} .\] 
Now let us consider the factor algebra $\f_1 \sim \g/ \z_1$. Its center is $C(\f_1) = C(\g/ \z_1)$. We define the {\it second
 center} of $\g$ to be 
\be
\z_2 = \z_1 \oplus C(\g/ \z_1).
\ee
Recursively we define higher centers as
\be
\z_{k+1} = \z_k \oplus C(\g/ \z_k).
\ee
For nilpotent Lie algebras the upper central series terminates, i.e. there exists $l$ such that
$\z_l = \g.$
We shall call these three series the {\it characteristic series} of the algebra $\g$. We shall use the notations $DS,CS$ and $US$ for
sets of integers denoting the dimensions of subalgebras in the derived, lower central and upper central series, respectively.

The {\it centralizer} $\g_\h$ of a given subalgebra $\h \subset \g$ in $\g$ is the set of all elements in $\g$ commuting with all
elements in $\h$, i.e.
\be
\g_\h = \{ x \in \g | [x,y]=0, \ \forall y \in \h \} .
\ee

A {\it derivation} $D$ of a given Lie algebra $\g$ is a linear map 
\[D: \ \g \rightarrow \g \]
such that for any pair $x,y$ of elements of $\g$ 
\be\label{deriv} 
D([x,y])=[D(x),y]+[x,D(y)]. 
\ee
If an element $z \in \g$ exists,  such that 
\[ D = {\rm ad}_{z}, \; \; {\rm i.e.} \; D(x)=[z,x], \ \forall x \in G \] ,
the derivation is called an {\it inner derivation}, any other one is an {\it outer derivation}.

\subsection{Solvable Lie algebras with a given nilradical}\label{introsan}

Any solvable Lie algebra $\s$ contains a unique maximal nilpotent ideal
$\n=NR(\s)$, the nilradical $\n$.
The dimension of the nilradical satisfies (\ref{dimnilrad}) \cite{Jac}.
We will assume that $\n$ is known. That is, in some basis $( e_1, \ldots, e_n )$ we know the Lie brackets
\be\label{nilkom}
[e_a,e_b] = N^c_{ab} e_c.
\ee
We wish to extend the nilpotent algebra $\n$ to all possible indecomposable solvable Lie algebras $\s$ having $\n$ as their
nilradical. Thus, we add further elements $f_1,\ldots,f_p$ to the basis $( e_1, \ldots, e_n )$ which together 
form a basis of $\s$. The derived algebra of a solvable Lie algebra is contained in the nilradical \cite{Jac}, i.e.
\be\label{ssinn}
[\s,\s] \subseteq \n.
\ee
It follows that the Lie brackets on $\s$ satisfy
\bea\label{Agam1}
[f_i,e_a] & = & (A_i)^b_{a} e_b, \; 1 \leq i \leq p, \  1 \leq a \leq n, \\ 
\label{Agam2} [f_i,f_j] & = & \gamma^a_{ij} e_a, \; 1 \leq i,j \leq p.
\eea

The matrix elements of the matrices $A_i$ must satisfy certain linear relations following from the Jacobi relations between
the elements $(f_i,e_a,e_b)$. The Jacobi identities between the triples  $(f_i,f_j,e_a)$ will provide 
linear expressions for the structure constants $\gamma^a_{ij}$ in terms of the matrix elements of the commutators of the 
matrices $A_i$ and $A_j$.

Since $\n$ is the maximal nilpotent ideal of $\s$, the matrices $A_i$ must satisfy another condition; no nontrivial 
linear combination of them is a nilpotent matrix, i.e. they are {\it linearly nil--independent}.

Let us now consider the adjoint representation of $\s$, restrict it to the nilradical $\n$ and find 
${\rm ad}|_\n (f_k)$. 
It follows from the Jacobi identities that ${\rm ad}|_\n (f_k)$ is 
a derivation of $\n$. In other words, finding all sets of matrices $A_i$ in (\ref{Agam1}) satisfying the Jacobi identities
is equivalent to finding all sets of outer nil--independent derivations of $\n$
\be
 D^1={\rm ad}|_\n (f_1),\ldots,D^p={\rm ad}|_\n (f_p).
\ee
Furthermore, in view of (\ref{ssinn}), the commutators $[D^j,D^k]$ must be inner derivations of $\n$. This requirement 
determines the structure constants $\gamma^a_{ij}$, i.e. the Lie brackets (\ref{Agam2}), up to elements in the center $C(\n)$ of $\n$.

Different sets of derivations may correspond to isomorphic Lie algebras, so redundancies must be eliminated. In terms of the Lie
brackets (\ref{Agam1}) and (\ref{Agam2}) it means that the matrices $A_i$ and constants $\gamma^a_{ij}$ must be classified into 
equivalence classes and a representative of each class must be chosen. Equivalence is considered under the following transformations
\be
f_i \rightarrow \tilde f_i = \rho_{ij} f_j + \sigma_{ia} e_a, \; \; e_a \rightarrow \tilde e_a = R_{ab} e_b
\ee
where $\rho$ is an invertible $p \times p$ matrix, $\sigma$ is a $p \times n$ matrix and 
the invertible $n \times n$ matrix $R$ must be chosen so that the Lie brackets (\ref{nilkom}) are preserved.

\subsection{A type of indecomposable nilpotent Lie algebras}

Any nilpotent Lie algebra $\n$ will contain a maximal Abelian subalgebra $\a$, not necessarily unique. We have \cite{Morozov}
\be
\frac{1}{2} (\sqrt{8n+1}-1) \leq \dim \a \leq \dim \n.
\ee
If $\dim \a = \dim \n$, then $\n=\a$ is Abelian. The case that we are interested in is the next closest to Abelian, namely 
$\dim \n=n$, $\dim \a = n-1$. Let us choose a basis $( e_1, \ldots, e_{n-1}, e_n ) $ of $\n$, where $ (e_1, \ldots, e_{n-1} ) $
is a basis of $\a$. The Lie brackets for $\n$ are
\be
[e_j,e_k]=0, \ 1 \leq j,k \leq n-1, \; \; [e_k,e_n] = \sum_{l=1}^{n-1} N_{kl} e_l, \; 1 \leq k \leq n-1.
\ee
The matrix $N \in \F^{(n-1) \times (n-1)}$  must be a nilpotent matrix, otherwise the algebra $\n$ will
 not be nilpotent. Elements of the center $C(\n)$ will corespond to the kernel ${\rm Ker} \ N$ of the matrix $N$. Elements 
of the derived algebra $\n^{(1)}$ will correspond to the image ${\rm Im} \ N$ of the matrix $N$. In order for the algebra $\n$
to be indecomposable, we must have
\be\label{kerinim}
{\rm Ker} \ N \subseteq {\rm Im} \ N.
\ee
Performing a change of basis  within the Abelian algebra $\a$, we  can transform the matrix $N$ to its Jordan canonical form. 
This can be one indecomposable nilpotent block, or several blocks. The condition (\ref{kerinim}) forbids the presence of 
one--dimensional blocks. There are as many mutually nonisomorphic algebras as there are nonequivalent partitions of $n-1$ into sums 
of positive integers satisfying
\be
n-1=n_1+n_2+\ldots+n_l, \; n_i \geq n_{i-1}, \; n_i \neq 1.
\ee
We shall denote the corresponding Lie algebras $\n_{n,k}$ where $n=\dim \n_{n,k}$ and $k$ enumerates the different isomorphy classes
for given $n$, i.e. the number of allowed partitions of $n-1$.

The rest of this article will be devoted to the algebras $\n_{n,1}$ with the matrix $N_{n,1}$ given by one indecomposable Jordan block. 
We shall find all extensions of these algebras to solvable Lie algebras with nilradical $\n_{n,1}$. The Lie brackets for $\n_{n,1}$ 
were already given in Equation (\ref{nla}).

\section{Classification of solvable Lie algebras with the nilradical $\n_{n,1}$}\label{cslawn}

\subsection{Nilpotent algebra $\n_{n,1}$}

The Lie algebra $\n_{n,1}$ is defined by the Lie brackets (\ref{nla}) of the Introduction.
We shall consider $n \geq 4$.
The dimensions of the subalgebras in the characteristic series are
\be
 DS=[n,n-2,0], \; CS=[n,n-2,n-3,\ldots,1,0], \; US=[1,2,\ldots,n-2,n].
\ee
Its maximal abelian ideal $\a$ can be identified with the centralizer of the highest center
$\z_{n-2}= \sp \{ e_1, \ldots,e_{n-2} \}$, i.e. $\a =\sp \{ e_1, \ldots,e_{n-1} \}$. Hence $\a$ is unique.

In order to find all non--nilpotent derivations of $\n$ we assign to $D$ its matrix 
\[ D(e_a) = D_{ab} e_b \]
and evaluate the condition (\ref{deriv}) for basis elements $x=e_i, \ y=e_j$.
We find
\begin{itemize}
\item $ i=1<j<n : \ D_{1n} e_{j-1} = 0$,
\item $ 1<i<j<n : \  D_{in} e_{j-1} - D_{jn} e_{i-1} =0, $
\item $ i = 1,  j=n: \ \sum_{k=1}^{n-1} D_{1,k+1} e_{k} =0,$
\item $ 1 < i < j=n: \ \left( D_{i-1,i-1} - D_{ii} - D_{nn} \right) e_{i-1} + \sum_{k=1,k \neq i-1}^{n-1} 
\left( D_{i-1,k}- D_{i,k+1} \right) e_{k} =0,$
\end{itemize}
From the first and second equations we immediately get 
\[ D_{in} =0, \ 1 \leq i < n ,\]
from the third
\[ D_{1k} =0, \ 1<k .\]
In the last one the coefficients of linearly independent elements in the sum must be zero, therefore considering 
\[ ( D_{i-1,k}- D_{i,k+1} ) =0, \ k \neq i-1 \]
we by induction obtain
\be
 D_{ij} = 0,\   D_{ji} = D_{j-1,i-1} = D_{j-i+1,1} ,\ i<j .
\ee
The remaining recursion relations 
\[ D_{i-1,i-1} - D_{ii} - D_{nn} =0 \]
can be most easily solved from the ``lower right corner'', denoting 
\[ D_{nn} = \alpha, \; D_{n-1,n-1} = \beta \]
we find 
\be\label{gderiv}
D_{ii} =  (n-i-1) \alpha + \beta, \; 1\leq i \leq n-1 . 
\ee
Thus we have solved the derivation property (\ref{deriv}) for all basis elements of $\n$. Finding the inner derivations is elementary 
and we may write

\begin{lemma}\label{genderiv}
The algebra of derivations of the nilpotent algebra $\n_{n,1}$ is expressible in the standard basis 
$(e_1,\ldots,e_n)$ of $\n_{n,1}$ as the algebra of lower triangular matrices $D$ whose elements satisfy
\[ D_{ii} =  (n-i+1) D_{nn} + D_{n-1,n-1}, \; 1\leq i \leq n-1, \]
\[ D_{ji} = D_{j-i+1,1} ,\ i<j .\]
Its subalgebra of inner derivations is 
\[ \sp \{ {\rm ad}_{e_2}, \ldots ,{\rm ad}_{e_n} \}, \]
where
\[ ( {\rm ad}_{e_j} )_{ab} = \delta_{an} \delta_{b,j-1}, \ \forall j \leq n-1, \; ( {\rm ad}_{e_n} )_{ab} = -  \delta_{a,b+1}. \]
\end{lemma}

\subsection{Construction of solvable Lie algebras with nilradical $\n_{n,1}$}\label{csla}

As was explained in Subsection \ref{introsan}, to find all solvable Lie algebras with nilradical $\n_{n,1}$  we must find all nonequivalent
nil--independent sets
$ \{ D^1,\ldots, D^p \} $ of derivations $\n_{n,1}$. The equivalence is generated by the following transformations:
\begin{enumerate}
\item We may add  any inner derivation to $D^k$.
\item We may perform a change of basis in $\n$ such that the Lie brackets are not changed. Since any such change must inter alia 
preserve the lower central series and the subalgebra $\a$, the matrix of such transformations must be lower triagonal. The 
preservation of Lie brackets then imposes certain further relations. We may decompose
any such transformation
into a composition of scaling 
\be\label{scal}
e_n \rightarrow \tilde e_n = \omega e_n, 
\ \ e_{k} \rightarrow \tilde e_{k} = \tau \omega^{n-k-1} e_{k}, \ k \leq n-1
\ee
and the transformation
\bea
\nn e_k \rightarrow \tilde e_k  & = & e_k + \sum_{j=1}^{k-1} u_{k-j} e_j, k \leq n-1, \ u_{1},\ldots,u_{n-2} \in \F, \\
\label{odtr} e_n \rightarrow \tilde e_n  & = & e_n + \sum_{j=1}^{n} v_j e_j, \  v_{1},\ldots,v_{n-1} \in \F.
\eea
(To prove that (\ref{odtr}) gives a general form of such a transformation it is sufficient to consider 
$[\tilde e_k,e_n]=[\tilde e_k,\tilde e_n]=\tilde e_{k-1}$ and to use induction on $k$.)
The scaling (\ref{scal}) acts on $D^k$ as 
\[ D^k \rightarrow S D^k S^{-1} \]
where 
$ S={\rm diag} ( \tau \omega^{n-2}, \tau \omega^{n-3}, \ldots, \tau, \omega ) $,
the transformation (\ref{odtr}) as 
\[ D^k \rightarrow U D^k U^{-1} \]
where
\[ U = \left( \ba{ccccccc}
1 & 0 &  0 &  0 & \ldots & 0 \\
u_{1} & 1 & 0 & 0  & \ldots & 0 \\
u_{2} & u_{1} & 1 &  0 & \ldots   & 0 \\
 & \ddots & \ddots & \ddots &  & \\
u_{n-2} & \ldots & u_{2}   & u_1 & 1 & 0 \\
v_1 &   \ldots & v_{n-3} & v_{n-2}  & v_{n-1} & 1 \ea \right).
\]
\item We can change the basis in the space $ \sp \{ D^1,\ldots, D^p \}$.
\end{enumerate}

By adding inner derivations we can transform all $D^k$ into the form
\be\label{canonD}
D^k = \left( {\ba{ccccccc} 
d^k_1 & 0 & 0 & 0 & \ldots   & 0 \\
0& d^k_2 & 0 &  0 & \ldots  & 0 \\
a_{3}^k & 0 & d^k_3 & 0 & \ldots & 0 \\
 & \ddots & \ddots &\ddots &    &  \\
a_{n-1}^k & \ldots & a_{3}^k & 0 & \beta^k & 0 \\
0 &  \ldots & 0 & 0 & a_{n}^k & \alpha^k 
\ea} \right),
\ee 
where 
\[d^k_1=(n-2) \alpha^k + \beta^k, \ldots ,d^k_j=(n-1-j) \alpha^k + \beta^k,\ldots,d^k_{n-2}=\alpha^k + \beta^k .\]
We shall assume that $D^k$ are always brought to this form.

We see that the number of nil--independent elements 
$p$ can be at most $2$ since a set of three or more derivations of the form (\ref{canonD}) cannot be 
 linearly nil--independent.

\begin{description}
\item[Case 1:] $p=1$ The entire
 structure of the associated solvable Lie algebra is encoded in the matrix $D$.
The Lie brackets of the non--nilpotent element $f$ with nilpotent elements are given by
\[ [f,e_k]= D(e_k) = D_{kl} e_l.\]
 We shall divide our investigation into subcases determined by values of the parameters 
$\alpha,\beta$, at least one of which must be nonzero.
\begin{enumerate}
\item $\alpha \neq 0$ \\
We rescale $D$ to put $\alpha=1$. Then by a change of basis (\ref{odtr}) in $\n$
\[ \tilde e_k = e_k - \frac{1}{l-1} a_l e_{k-l+1},\ l \leq k \leq n-1,\]
\[ \tilde e_k = e_k, \ 1 \leq k \leq l-1 \]
we put to zero first $a_3$, then $a_4$ etc. up to $a_{n-1}$. From now on we assume 
that $a_k=0, k\leq n-1$. If $\beta \neq 1 (=\alpha)$ then a further change of basis
\be\label{cb1}
 \tilde e_n = e_n - \frac{a_{n}}{\beta-1} e_{n-1} 
\ee 
turns $a_n$ into zero and the matrix $D$ is diagonal
\be\label{diagD}
 D = {\rm diag}( n-2 + \beta, n-3 + \beta, \ldots, \beta, 1 ).
\ee

If $\beta=1$ then $a_n$ cannot be removed. The only remaining transformation is scaling 
(\ref{scal}) which allows us to scale any nonzero $a_n$ to 1. Therefore we find in addition to 
(\ref{diagD}) with $\beta=1$ another possibility, namely
\be\label{ndD1}
 D = \left( {\ba{ccccc} 
n-1 & 0 &\ldots &  0  & 0 \\
0 & n-2 &\ldots &  0  & 0 \\
 & & \ddots &  &    \\
0 & 0 & \ldots & 1  &  0 \\
0 & 0 & \ldots & 1 & 1 
\ea} \right) 
\ee

\item $\alpha=0$ \\
We rescale $D$ to put $\beta=1$. We use (\ref{odtr}) to change $e_n$
\[ \tilde e_n= e_n - a_{n} e_{n-1}\]
and transform $a_{n}$ into $a_{n}=0$. If $D$ is diagonal, it cannot be further simplified.
Let us assume that $D$ is not diagonal. For $\alpha=0$ the matrix $D$ is invariant with 
respect to transformations (\ref{odtr}) preserving $a_{n}=0$, i.e. the parameters 
$a_k$ cannot be removed. The only transformation we still have at our disposal is the scaling (\ref{scal})
which allows us to scale one chosen nonzero $a_{k}$ to $1$ over the field $\C$. Over $\R$ one value $a_k$ can be scaled to $1$
if $k$ is even, or to $\pm 1 $ if $k$ is odd.
\end{enumerate}
\item[Case 2:] $p=2$ By taking linear combinations of $D^1,D^2$ we obtain
$ \alpha^1=1, \ \beta^1=0, \ \alpha^2=0, \ \beta^2=1.$
Further by a change of basis in $\n$ (\ref{odtr}) we take $D^1$ to its canonical form
\[ D^1= {\rm diag} (n-2,n-3,\ldots,2,1,0,1)\]
found for $p=1$.
In order to define a solvable Lie algebra $\g$ with nilradical $\n$, the two derivations $D^1,D^2$ must commute to 
an inner derivation
\[ [D^1,D^2] \in \sp \{ {\rm ad}_{e_2},\ldots,{\rm ad}_{e_n} \}. \]
Computing the commutator for the above given forms of $D^1,D^2$ (note that $D^1$ is diagonal) we immediately find that 
\[ a_{k}^2=0, \ 3 \leq k \leq n \]
must hold.

Therefore there is a single canonical form of $D^1,D^2$ 
\be
D^1  =  {\rm diag} (n-2,n-3,\ldots,2,1,0,1), \;\; D^2  =  {\rm diag} (1,1,\ldots,1,1,0).
\ee

The corresponding solvable Lie algebra is now almost specified, the Lie brackets of non--nilpotent elements $f_1,f_2$ being
\[ [f_1,e_k]=  (n-1-k) e_k, \ k<n, \ [f_1,e_n]=e_n,\]
\[ [f_2,e_k]= e_k, \ k<n, \ [f_2,e_n]=0.\]
It remains to fix the Lie bracket between $f_1,f_2$. Because $D^1$ and $D^2$ are commuting  matrices representing $f_1,f_2$ in 
the adjoint representation of $\g$ restricted to the ideal $\n$, the Lie bracket of $f_1,f_2$ must be in the kernel 
of the representation map, i.e. in the center of $\n$
\be\label{gamma}
 [f_1,f_2] = \gamma e_1 .
\ee
The transformation
\[ f_1 \rightarrow \tilde f_1 = f_1 + \gamma e_1 \]
takes $\gamma$ in Equation (\ref{gamma}) into $\gamma=0$
while leaving all other Lie brackets invariant.
We conclude that in the case $p=2$ the solvable Lie algebra with nilradical $\n_{n,1}$ is unique.
\end{description}

\subsection{Standard forms of solvable Lie algebras with nilradical $\n_{n,1}$}

The results obtained above can be summed up as theorems. We give them for algebras over the field $\F=\R$ or $\C$. 
We specify $\F=\R$, or $\C$ only when the two cases differ. In all cases we give the dimensions of the subalgebras in 
the characteristic series. These dimensions are basis independent and are very useful for identifying the Lie algebras.
The nilradical in all cases is $\n_{n,1}$ with the Lie brackets (\ref{nla}). We shall specify the action of the (standardized)
nonnilpotent elements $f$ or $f_1$ and $f_2$ on the nilradical (see Equation (\ref{Agam1})). 

In the theorems, ``solvable'' will always mean solvable, indecomposable, nonnilpotent.

\begin{veta}\label{th1}
Any solvable Lie algebra $\s$ with nilradical $\n_{n,1}$ will have dimension $\dim \s=n+1$, or $\dim \s=n+2$.
\end{veta}

\begin{veta}\label{th2}
Three types of solvable Lie algebras of dimension $\dim \s=n+1$ exist for any $n \geq 4$. They are represented by the following:
\begin{enumerate}
\item $A=A_1$ in Equation (\ref{Agam1}) diagonal
\be
[f,e_k]= \left( (n-k-1)\alpha+\beta \right) e_k, \ k \leq n-1, [f,e_n]=\alpha e_n.
\ee
The mutually nonisomorphic algebras of this type are
\bea
 \s_{n+1,1}(\beta): & & \alpha=1, \beta \in \F \backslash \{ 0,n-2 \},  \\
\nn DS  & = &  [n+1,n,n-2,0], \ CS=[n+1,n,n,\ldots], US=[0], \\
 \s_{n+1,2}: & & \alpha=1, \beta=0,  \\
\nn  DS  & = &  [n+1,n-1,n-3,0], \ CS=[n+1,n-1,n-1,\ldots], US=[0], \\
 \s_{n+1,3}: & & \alpha=1, \beta=2-n,  \\
\nn  DS  & = &  [n+1,n,n-2,0], \ CS=[n+1,n,n,\ldots], US=[1,1,\ldots], \\ 
 \s_{n+1,4}: & & \alpha=0, \beta=1,  \\
\nn  DS  & = &  [n+1,n-1,0], \ CS=[n+1,n-1,n-1,\ldots], US=[0]. 
\eea
\item $A=A_1$ in Equation (\ref{Agam1}) nondiagonal, its diagonal determined by $\alpha=\beta=1$. We have
\bea
\s_{n+1,5}: & &  [f,e_k]=(n-k) e_k, \ k \leq n-1, [f,e_n]=e_n+e_{n-1}, \\
\nn DS & = & [n+1,n,n-2,0], \ CS=[n+1,n,n,\ldots], \ US=[0]. 
\eea
\item $A=A_1$ in Equation (\ref{Agam1}) nondiagonal, its diagonal determined by $\alpha=0,\beta=1$. 
\bea
\nn \s_{n+1,6}(a_3,\ldots,a_{n-1}): & &  [f,e_k]=e_k + \sum_{l=1}^{k-2} a_{k-l+1} e_l, \ k \leq n-1, \\
 & & [f,e_n]=0, 
\eea
$a_j \in \F$, at least one $a_j$ satisfies $a_j \neq 0$. \\
Over $\C$: the first nonzero $a_j$ satisfies $a_j=1$.\\
Over $\R$: the first nonzero $a_j$ for even $j$ satisfies $a_j=1$. If all $a_j=0$ for $j$ even, then the first nonzero $a_j$ 
($j$ odd) satisfies $a_j= \pm 1$. We have
\bea
\nn DS  & = &  [n+1,n-1,0], \ CS=[n+1,n-1,n-1,\ldots], US=[0]. 
\eea
\end{enumerate}
\end{veta}

\begin{veta}\label{th3}
Precisely one class of solvable Lie algebras $\s_{n+2}$ of $\dim \s=n+2$ with nilradical $\n_{n,1}$ exists. It is represented by a basis 
$( e_1, \ldots, e_n,f_1,f_2 )$ and the Lie brackets involving $f_1$ and $f_2$ are
\bea
\nn [f_1,e_k] & = & (n-1-k)e_k, \ 1 \leq k \leq n-1, \ [f_1,e_n]=e_n, \\
  {[f_2,e_k]} & = & e_k, \ 1 \leq k \leq n-1, \ [f_2,e_n]=0, \ [f_1,f_2]=0.
\eea
For this algebra
\bea
DS  & = &  [n+2,n,n-2,0], \ CS=[n+2,n,n,\ldots], US=[0] .
\eea
\end{veta}

\section{Generalized Casimir invariants}\label{GCI}

\subsection{General method}

The term {\it Casimir operator}, or {\it Casimir invariant}, is usually reserved for elements of the center of 
the enveloping algebra of a Lie algebra $\g$ \cite{Casimir,Racah}. These operators are in one--to--one correspondence
with polynomial invariants characterizing orbits of the coadjoint representation of $\g$ \cite{Kirillov}. The search for invariants
of the coadjoint representation is algorithmic and amounts to solving a system of linear first order partial differential equations 
\cite{Abellanas, Abellanas1,CS,CS1,Perroud,PSW,RW,NW1,TW1}. Alternatively, global properties of the coadjoint representation 
can be used \cite{Perroud}. In general, solutions are not necessarily polynomials and we shall call the nonpolynomial solutions 
{\it generalized Casimir invariants}. For certain classes of Lie algebras, including semisimple Lie algebras, perfect 
Lie algebras, nilpotent Lie algebras, and more generally algebraic Lie algebras, all invariants of the coadjoint representation 
are functions of polynomial ones \cite{Abellanas, Abellanas1}. 

Casimir invariants are of primordial importance in physics. They represent such important quantities as angular momentum, elementary 
particle's mass and spin, Hamiltonians of various physical systems etc.

In the representation theory of solvable Lie algebras the invariants are not necessary polynomials, i.e. they can be genuinely 
generalized Casimir invariants. In addition to their importance in representation theory, they may occur in physics. Indeed, 
Hamiltonians and integrals of motion for classical integrable Hamiltonian systems are not necessarily polynomials in the momenta 
\cite{Hietarinta,Ramani}, though typically they are invariants of some group action.

In order to calculate the (generalized) Casimir invariants we consider some basis $(g_1,\ldots,g_n)$ of $\g$, in which 
the structure constants are $c^k_{ij}$.
A basis for the coadjoint representation is given by the first order differential operators
\be\label{doal}
 \hat G_k = g_b c^b_{ka} \frac{\pd}{\pd g_a}.
\ee
In Equation (\ref{doal}) the quantities $g_a$ are commuting independent variables.

The invariants of the coadjoint
representation, i.e. the generalized Casimir invariants, are solutions of the following system of partial differential equations
\be\label{casimir}
 \hat G_k I(g_1, \ldots,g_n)=0, \ k=1,\ldots,n .
\ee
The number of functionally independent solutions of the system (\ref{casimir}) is
\be
n_I=n-r(C)
\ee
where $C$ is the antisymmetric matrix
\be
C = \left( \ba{cccc} 0 & c^b_{12} g_b & \ldots & c^b_{1n} g_b \\
-c^b_{12} g_b & 0 & \ldots & c^b_{2n} g_b \\
\vdots & & & \vdots \\
-c^b_{1,n-1} g_b & \ldots & 0 & c^b_{n-1,n} g_b \\
-c^b_{1n} g_b & \ldots & -c^b_{n-1,n} g_b & 0 \ea \right)
\ee
and $r(C)$ is the generic rank of $C$. Since $C$ is antisymmetric, its rank is even. 
Hence $n_I$ has the same parity as $n$. 

Since the method of computation is generally known, we shall not present details and just give the results in 
the form of theorems. In all cases proofs consist of a direct calculation, i.e. solving Equations (\ref{casimir}).

\subsection{The generalized Casimir invariants}

The differential operators corresponding to the basis elements of $\n_{n,1}$ are
\bea
\hat E_1 & = & 0, \ \ \hat E_k  =  e_{k-1} \frac{\pd}{\pd e_n}, \; 1<k<n , \ \ \hat E_n  =  - \sum_{k=2}^{n-1} e_{k-1} \frac{\pd}{\pd e_k}.
\eea
The form of $\hat E_k, \ (1<k<n)$ implies that the invariants do not depend on $e_n$. Solving the equation 
$\hat E_n \ I(e_1,e_2,\ldots,e_{n-1})=0$  by the method of characteristics, we obtain the following result
\begin{veta}\label{th4}
The nilpotent Lie algebra $\n_{n,1}$ has $n-2$ functionally independent invariants. They can be chosen to be the following polynomials
\bea
\nn \xi_0 & = & e_1, \\
\label{nac} \xi_k & = & \frac{(-1)^k k}{(k+1)!} e_2^{k+1} + \sum_{j=0}^{k-1} (-1)^j \frac{e_2^j \ e_{k+2-j} \ e_1^{k-j}}{j!}, 
\; 1 \leq k \leq n-3.
\eea
\end{veta}

Let us now consider the $(n+1)$ dimensional solvable Lie algebras of Theorem \ref{th2}. The operators $\hat E_i$ 
representing $\n_{n,1}$ will each contain an additional term involving a derivative with respect to $f$. However, 
from the form of these operators we see that the invariants cannot depend on $f$. Moreover, they can only depend 
on the invariants (\ref{nac}) of $\n_{n,1}$. To find the invariants of the algebras $\s_{n+1,k}$ we must represent the element
$f\in \s_{n+1,k}$ by the appropriate ``truncated'' differential operator $\hat F_{T}$ (by ``truncated'' we mean that 
we keep only the part acting on $e_1,\ldots,e_{n-1}$). We must then solve the equation 
\be\label{FTI}
\hat F_{T} I(\xi_0,\ldots,\xi_{n-3})=0.
\ee
For the algebras $\s_{n+1,1},\ldots,\s_{n+1,5}$ of Theorem \ref{th2}, we have
\be
\label{n11do}
 \hat F_T = \sum_{k=1}^{n-1} ((n-1-k)\alpha+\beta) e_k \frac{\pd}{\pd e_k}
\ee
with $\alpha$ and $\beta$ as in Theorem \ref{th2}. For $\s_{n+1,6}(a_3,\ldots,a_{n-1})$ we have
\be
\label{odacf} \hat F_T  =  e_1 \frac{\pd}{\pd e_1} + e_2 \frac{\pd}{\pd e_2}+ \sum_{l=1}^{n-3} \left( e_{l+2} + \sum_{j=1}^{l}
a_{l+3-j} e_{j} \right) \frac{\pd}{\pd e_{l+2}}.
\ee
Solving Equation (\ref{FTI}) in each case we obtain the following result.

\begin{veta}\label{th5}
The algebras $\s_{n+1,1}(\beta),\ldots,\s_{n+1,5}$ have $n-3$ invariants each. Their form is
\begin{enumerate}
\item $\s_{n+1,1}(\beta)$, $\s_{n+1,2}$ and $\s_{n+1,5}$
\be\label{sn111215inv}
\chi_k = \frac{\xi_k}{\xi_0^{(k+1)\frac{n-3+\beta}{n-2+\beta}}}, \ 1 \leq k \leq n-3.
\ee
For $\s_{n+1,2}$ and $\s_{n+1,5}$ we have $\beta=0$ and $\beta=1$, respectively in Equation (\ref{sn111215inv}).
\item $\s_{n+1,3}$
\be
\chi_1 = \xi_0, \  \chi_k = \frac{\xi^2_{k}}{\xi_1^{k+1}}, \  2 \leq k \leq n-3.
\ee
\item $\s_{n+1,4}$
\be
\chi_k = \frac{\xi_{k}}{\xi_0^{k+1}}, \  1 \leq k \leq n-3.
\ee
\item $\s_{n+1,6}(a_3,\ldots,a_{n-1})$
\bea
 \chi_k & = & \sum_{m=0}^{[\frac{k+1}{2}]} (-1)^m \frac{(\ln \xi_0)^m}{m!} \left( \sum_{i_1+\ldots+i_m=k-2m+1} a_{i_1+3}  
a_{i_2+3}\ldots a_{i_m+3} \right. \\
\nn & + & \left. \sum_{j+ i_1+\ldots+i_m=k-2m-1} \frac{\xi_{j+1}}{\xi_0^{j+2}} \ a_{i_1+3}  a_{i_2+3}\ldots a_{i_m+3} \right),
\ 1 \leq k \leq n-3
\eea
The summation indices take the values $0 \leq j,i_1,\ldots,i_m \leq k+1$.
\end{enumerate}
\end{veta}
Finally, let us consider the $(n+2)$ dimensional algebra $\s_{n+2}$. The invariants can again depend only on $\xi_0,\ldots,\xi_{n-3}$.
We have two additional truncated differential operators, namely
\be
 \hat F_{1T}  =  \sum_{k=1}^{n-1} (n-1-k) e_k \frac{\pd}{\pd e_k}, \; \;  \hat F_{2T}  =  \sum_{k=1}^{n-1} e_k \frac{\pd}{\pd e_k}. 
\ee
Imposing two equations of the form (\ref{FTI}) we obtain the following result
\begin{veta}\label{th6}
The Lie algebra $\s_{n+2}$ of Theorem \ref{th3} has $n-4$ functionally independent invariants that can be chosen to be 
\be
\chi_k = \frac{\xi_{k+1}}{\xi_1^{\frac{k+2}{2}}}, \; 1 \leq k \leq n-4  .
\ee
\end{veta}

We see that for the algebra $\s_{n+1,6}(a_3,\ldots,a_{n-1})$ the invariants involve powers of the logarithm $\ln \xi_0$.
In all other cases we obtain sets of ratios of powers of $\xi_k$.

A specific class of solvable Lie algebras, namely rigid ones, was considered by R. Campoamor-Stursberg \cite{CS1}, who calculated 
their generalized Casimir invariants for dimensions up to $N=8$ inclusively. Our algebras $\s_{n+2}$ fall into this category 
(with $N=n+2$). Our results for $n \leq 6$ agree with those of \cite{CS1}.

\section{Conclusions}

The main results of this article are summed up in Theorems \ref{th1},\ref{th2} and \ref{th3} of Section \ref{cslawn}
and Theorems \ref{th4},\ref{th5} and \ref{th6} of Section \ref{GCI}. 

The results on the structure of indecomposable solvable Lie algebras with the nilradical $\n_{n,1}$ are quite simple and 
it is interesting to compare them with results for other nilradicals. This comparison is performed in Table \ref{tb1}.
\begin{table}[tb] 
\caption{\label{tb1} Number of linearly nil--independent elements that can be added to the nilradical.} 
\begin{indented} 
\item[]\begin{tabular}{@{}lll} 
\br Nilradical $\n$ & $\dim \ \n$ & $p_{{\rm max}}$ \\ 
\mr 
$\n_{n,1}$ & $n \geq 4$  & 2 \\
$\a_n$ & $n \geq 3$  & $n-1$ \\
$\a_1$ & $n=1$ & 1 \\
$\a_2$ & $n=2$ & $1 \ (\C)$, $2 \ (\R)$ \\
$\h(N)$ & $n=2N+1, \ N \geq 1$ & $N+1$ \\
$\t(N)$ & $n=\frac{(N-1)N}{2}, \ N \geq 2$ & $N-1$ \\
\br 
\end{tabular} 
\end{indented} 
\end{table}
There $\a_n$ denotes an $n$ dimensional Abelian Lie algebra, $\h(N)$ a Heisenberg algebra in an $N$ dimensional space and $\t(N)$
is the subalgebra of strictly upper triangular matrices of $\sl(N,\F)$. In the third column $p_{{\rm max}}$ is the maximal number 
of nonnilpotent elements we can add in order to obtain an indecomposable solvable Lie algebra. Notice that $p_{{\rm max}}$
is independent of the dimension of the nilradical only for $\n_{n,1}$.

The results on generalized Casimir invariants are also quite simple. For the nilradical $\n_{n,1}$ the number of invariants is 
$n_I=n-2-p$, where $p$ is the number of nonnilpotent elements (i.e. $p=1$ or $p=2$). By comparison, for the Abelian nilradical 
$\a_{n}$, the number of invariants is $n_I=n-p$. In both cases the invariants depend only on elements of the nilradical and 
can be polynomials, ratios of powers of polynomials, or may involve logarithms.

Finally, a few words about applications of the Lie algebras obtained above. The algebras $\s_{n+1,j}$ of Theorem \ref{th2}, for $n=4$,
appear in Petrov's classification \cite{Petrov} of gravitational fields admitting groups of motion (isometry groups) of dimension 5.
The fact that we have a complete list of all such Lie algebras for arbitrary $n$ would enable us to construct the corresponding 
invariant Riemann, or pseudo--Riemann metrics in spaces of arbitrary dimension. An invariant metric then makes it possible to write 
invariant classical, or quantum integrable systems in such spaces and to investigate the separation of variables in Hamilton--Jacobi 
and Schr\"odinger equations.

\ack The research of Pavel Winternitz was partly supported by a research grant from NSERC of Canada. Libor \v Snobl acknowledges
a postdoctoral fellowship awarded by the Laboratory of Mathematical Physics of the CRM, Universit\'e de Montr\'eal.

\end{document}